\documentclass[11pt]{article}
\usepackage{jheppub}
\usepackage{mathrsfs} 
\usepackage{setspace}
\usepackage{amsmath, amsfonts}
\numberwithin{equation}{section}  
\usepackage[utf8]{inputenc}
\usepackage{graphicx}
\usepackage{ifthen}
\usepackage{youngtab}

\makeatletter
\def\@fpheader{\relax}
\makeatother

\DeclareMathOperator{\Tr}{Tr}
\DeclareMathAlphabet{\mathbbold}{U}{bbold}{m}{n} 
\DeclareMathOperator*{\genhyp}{F}
\newcommand{\sun}[1][]{SU(N)_{#1}}

\newcommand{\vertex}[2]{V_{#2}(x_{#1})}
\newcommand{\entao}{\quad\Rightarrow\quad}
\newcommand{\Dpfrac}[3][]{\frac{\partial^{#1} #2}{\partial {#3}^{#1}}}
\newcommand{\del}[1][z]{\partial_{#1}}

\title{Wilson Loop Invariants from $W_N$ Conformal Blocks}

\author[]{Oleg Alekseev}\emailAdd{alekseev@itp.ac.ru}
\author[]{and Fábio Novaes}\emailAdd{fabio.nsantos@gmail.com}

\affiliation[]{International Institute of Physics, Federal University of Rio Grande do Norte,
Av. Odilon Gomes de Lima 1722, Capim Macio, Natal-RN 59078-400, Brazil}

\abstract{ Knot and link polynomials are topological invariants
  calculated from the expectation value of loop operators in
  topological field theories. In 3D Chern-Simons theory, these
  invariants can be found from crossing and braiding matrices of
  four-point conformal blocks of the boundary 2D CFT. We calculate
  crossing and braiding matrices for $W_N$ conformal blocks with one
  component in the fundamental representation and another component in
  a rectangular representation of $\sun$, which can be used to obtain
  HOMFLY knot and link invariants for these cases. We also discuss how
  our approach can be generalized to invariants in
  higher-representations of $W_{N}$ algebra.}  

\preprint{\today}

\begin{document}

\maketitle{}

\section{Introduction}\label{sec:resumocientifico}



In the late 80's, the important connection between quantum field
theory (QFT) and Jones Polynomials \cite{Jones1987} was uncovered by
Edward Witten \cite{Witten1989a} showing that knot theory is deeply
connected to topological QFTs. The particular example studied by
Witten was $SU(N)$ Chern-Simons theory (CS) in a 3D compact manifold
$M$. Non-trivial states in CS are topological as there are no
dynamical degrees of freedom in this theory. Therefore, Wilson loops
are natural elementary states of CS. To construct a non-trivial state
in this sense, we must create knotted loops or link several loops
together. The expectation value of these composite objects gives us
certain topological invariants called \emph{knot (or link)
  polynomials}. For the $SU(2)$ case, Witten has shown that one
obtains Jones polynomials from the Wilson loop expectation
values. This relation between Chern-Simons and knot theory is an
important example of integrability in quantum field theories, which
serves as a tool to organize and construct physical theories.

The explicit construction of these invariants starts with a partition
of $M$ into two manifolds with a boundary. To each boundary is
associated a WZNW theory whose Hilbert space is the space of conformal
blocks with Wilson lines as punctures on the boundary. Braiding
matrices of the boundary CFT can then be used to construct the
original knot or link in $M$
\cite{Witten1989a,Kaul1992,Kaul1993,RamaDevi1993}. Here we call this
the \emph{crossing-matrix method}. This approach can be related to
quantum groups, as the crossing transformations of conformal blocks
are directly related to $SU(N)_{q}$ quantum Racah
matrices~\cite{alvarez1990duality,reshetikhin1990,Mironov2011,Mironov2012}.
For the most recent and broad discussion of this method, see
\cite{Mironov2015}. When combined with the evolution method
\cite{Mironov2013g} and cabling procedure \cite{Anokhina:2013ica}, the
crossing matrices can be extrapolated to give explicit formulas for
many families of knots and links \cite{Galakhov2015}. This combination
of techniques gives not only Jones polynomials $J_{R}^{C}(q)$
\cite{D.Galakhov:2015aa} but also can be uplifted to calculate HOMFLY
polynomials $H_{R}^{C}(q,A)$ \cite{freyd1985new,Mironov2014} and
\emph{superpolynomials} $P^{C}_{R}(q,A,t)$. Superpolynomials appeared
in physics in the connection between topological string theories,
M-theory and Chern-Simons
\cite{Ooguri2000,Gukov2005,dunfield2006superpolynomial}. 
Understanding better how
HOMFLY polynomials come about in $W_{N}$ models might also shed some
light on the nature of superpolynomials.  The crossing-matrix method
has also been used to construct loop operators of $\mathcal{N}=2$
gauge theories via its AGT relation with Liouville
\cite{Alday2010a,Drukker2010} and Toda field theory
\cite{Wyllard2009a,Passerini2010,Drukker2011}, so the study of $W_{N}$
conformal blocks is also interesting for this AGT approach.

Going back to the relation between knots and conformal blocks, the
crossing-matrix method is well-understood in the Virasoro case
($W_{2}$ algebra), but has not been explicitly developed before for
$W_{N}$ algebras to the authors knowledge. Here we develop the
crossing-matrix approach for $W_{N}$ minimal models directly from the
CFT point of view in detail. Evidence has been put forward in
\cite{Ramadevi1994a} that knot and link invariants in $W_{N}$ models
should factorize in terms of $SU(N)_{q}$ invariants, but the crossing
matrices have not been calculated. Much more is known about
$SU(N)_{q}$ Racah matrices and topological invariants constructed with
it \cite{Zodinmawia:2011oya,Nawata:2013qpa}, including representations
with non-trivial multiplicities \cite{Gu2014b}. Therefore, as proposed
in \cite{Ramadevi1994a}, we expect that $W_{N}$ invariants should
reduce to $SU(N)_{q}$ invariants and indeed that is what we find in
the cases studied below. However, we still have limited information
about higher-representations and it is not clear if the
crossing-matrices will factorize in general for $W_{N}$ correlators.

Four-point Virasoro conformal blocks need only one completely
degenerate field to obey a hypergeometric differential equation, the
BPZ equation \cite{Belavin1984}. However, for higher $W_{N}$ algebras
($N > 2$), we need one more constraint to find a differential equation
and to obtain explicit crossing $S$ and braiding $T$ matrices
\cite{Bowcock1994a,Fateev:2005aa}. If we also set some other field to
be semi-degenerate, the conformal blocks obey a generalized
hypergeometric equation. In this note, we construct $S$ and $T$
matrices with two fields in the fundamental and anti-fundamental
representations of $SU(N)$ and the other two in a rectangular
representation and its conjugate. These cases are somewhat degenerate
with respect to higher-representations of $W_{N}$ primary fields, as
the dimension of the space of conformal blocks is two-dimensional, but
are the first step to obtain more general S-matrices for
higher-representations \cite{D.Galakhov:2014aa,D.Galakhov:2015aa}.  In
our particular case, the generalized hypergeometric equation reduces
to a Gauss hypergeometric equation, for which the connection formulas
are explicitly known and, thus, the crossing matrices.



In section \ref{sec:wilson-loops-chern}, we revise the relation
between Wilson loop operators in 3D Chern-Simons, knot invariants and
conformal blocks. In section \ref{sec:toda-field-theory}, we set up our notation by reviewing
how to obtain knot invariants in Virasoro
models. 
In section \ref{sec:knot-invariants}, we discuss how to calculate knot
and link invariants from $W_{N}$ conformal blocks. Finally, in section
\ref{sec:conclusions} we present our conclusions and discuss further
developments.

\section{Knot and Link Invariants from Conformal Blocks}
\label{sec:wilson-loops-chern}

\begin{figure}
\centering
\includegraphics[width=\textwidth]{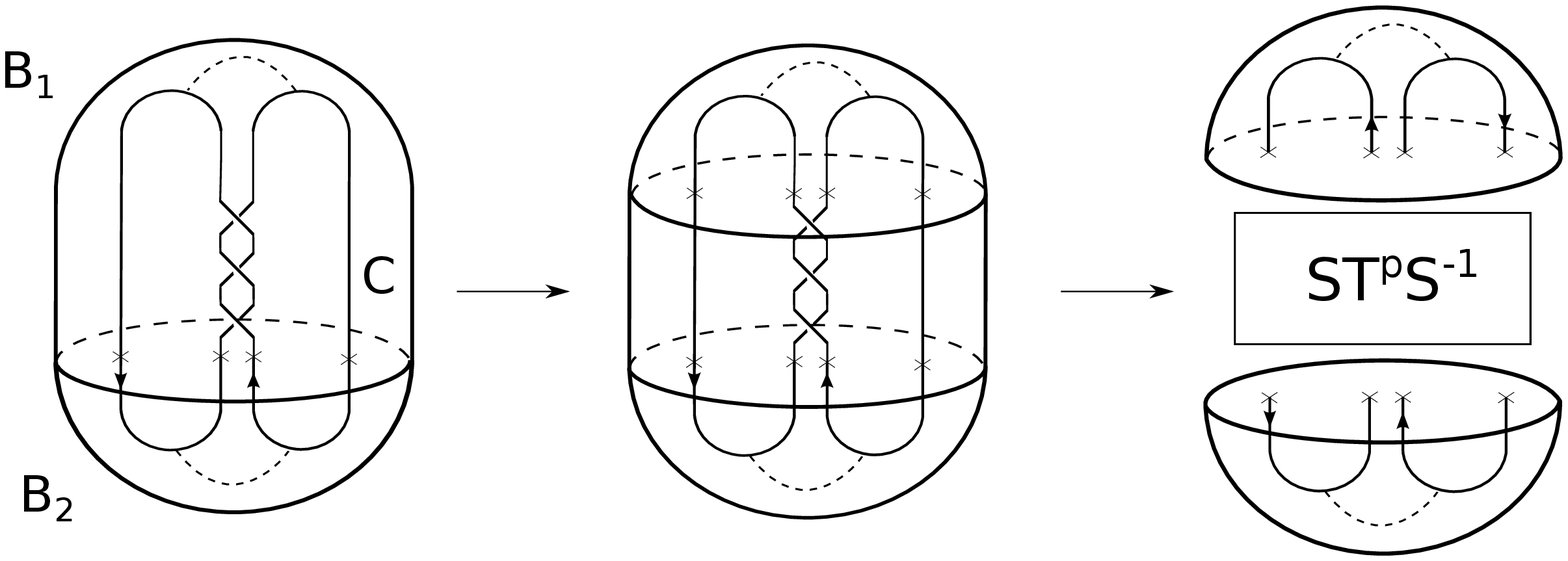}
\caption{Splitting a two-bridge knot $C$ into an inner product of
  conformal blocks. The brading and fusion matrices allow for a
  reconstruction of $C$ from basic conformal blocks with zero weight
  intermediate states.}
\label{fig:innerprod} 
\end{figure}

Following Witten's construction \cite{Witten1989a}, we are interested
in calculating the expectation value of non-trivial Wilson loops
forming a knot or link $C$ embedded in a closed three-dimensional
manifold $M$ in Chern-Simons theory. For simplicity, we take $M$
isomorphic to a 3-sphere. We can then cut $C$ into two bounded parts,
$B_{1}$ and $B_{2}$, by slicing $M$ with a 2-dimensional surface (see
fig.~\ref{fig:innerprod}). To each $B_{k}$ we relate a state
$\psi_{k}$ of the WZNW CFT defined on its boundary $\partial B_{k}$.
In this interpretation, a Wilson loop invariant is given by the inner
product between these two states
\begin{equation}
  Z_{\mathrm{R}}(C) = \left\langle \Tr_{\mathrm{R}} \mathcal{P}\exp\left(\oint_{C} A \right) \right\rangle_{CS} = \langle\psi_{1}|\psi_{2}\rangle.
\end{equation}
The Hilbert space $\mathcal{H}_{k}$ of each $B_{k}$ is isomorphic to
the space of conformal blocks of the boundary CFT. These blocks have
extra proportionality parameters coming from the braiding and crossing
operations to build up $C$, as explained below. Here and in the rest
of the paper we restrict our attention to invariants build up from
four-point conformal blocks, also called \emph{two-bridge} states.

\begin{figure}
  \centering
  \includegraphics[width=\textwidth]{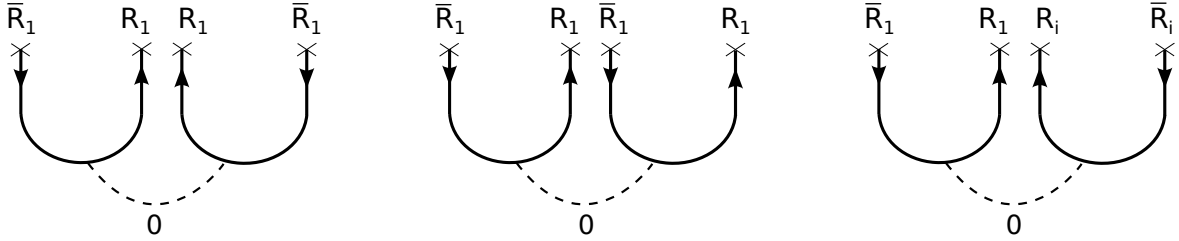}
  \caption{Different possibilities of basic conformal blocks. From
    left to right: paralell, antiparallel and mixed case. }
\label{fig:cblocks}
\end{figure}

The two-bridge knot invariants can be constructed via braiding and
closure of a $j$-channel conformal block $\mathcal{F}^{j}_{0}$ with
zero weight\footnote{As the total charge of a closed manifold must be
  zero.} intermediate state, where the index $j$ represents either the
s, t or u-channel.  Each puncture represented in
fig.~\ref{fig:innerprod} corresponds to a field insertion of the
conformal block in some representation $\mathrm{R}$ of $SU(N)$, as shown in
fig.~\ref{fig:cblocks}. We fuse the relevant fields via a crossing
matrix $S$ and then braid several times with a diagonal half-monodromy
matrix\footnote{These matrices are also called fusion matrix $F$ and
  braiding matrix $B$ in the literature. $S$ and $T$ usually refers to
  modular transformations on the torus.}  $T$. General S and T
matrices depend on the field representations
\begin{equation}
  S_{i_{1}i_{2}}
  \begin{bmatrix}
    \mathrm{R}_{2} & \mathrm{R}_{3}\\
    \mathrm{R}_{1} & \mathrm{R}_{4}
  \end{bmatrix},\quad T_{j_{1}j_{2}}[\mathrm{R}_{1}\; \mathrm{R}_{2}],
\end{equation}
with the internal indices being labeled by the result of the fusion of
appropriate representations, i.e.,
$i_{1},i_{2}\in (\mathrm{R}_{1}\otimes \mathrm{R}_{2})\cap (\mathrm{R}_{3}\otimes \mathrm{R}_{4})$ and
$j_{1},j_{2} \in [\mathrm{R}_{k}]$, where $[\mathrm{R}_{k}]$ represents the intermediate
states in the appropriate channel. Non-trivial multiplicities might
also appear in the fusion rules of certain fields but we do not
consider those here. In the following, we omit the matrix dependence
on representations.

Back to figure~\ref{fig:cblocks}, lines going up in representation
$\mathrm{R}_{j}$ must close with lines going down in the conjugate
representation $\bar{\mathrm{R}}_{j}$ after the braiding evolution in
the last step of fig.~\ref{fig:innerprod}. The first two cases in
fig.~\ref{fig:cblocks} have parallel and anti-parallel fusing strands,
respectively. The third case has one of the bridges in a different
representation. Only in the first two cases we can close the strands
to form a knot\footnote{In the second case, we can have twist knots
  \cite{D.Galakhov:2015aa}.} or a link and in the third case we have
only links. In this paper, we are going to consider the first two
cases with $\mathrm{R}_{1}$ in the fundamental representation of
$SU(N)$ and the third case with $\mathrm{R}_{i}$ in a rectangular
representation of $\sun$.

For different sequences made up of $S$ and $T$ matrices, we can
construct several types of knots
\cite{D.Galakhov:2015aa,Mironov2015}. The simplest examples of
two-bridge invariants are described by the following formula
\begin{equation}
  Z^{j,p}_{\mathrm{R}}(C)\equiv  \langle \mathcal{F}^{j}_{0}|S T^{p}
  S^{-1}|\mathcal{F}^{j}_{0}\rangle = (S T^{p}
  S^{-1})_{00}, \quad p\in\mathbb{Z}^{+},
\end{equation}
where the last equality represents the singlet diagonal component of
the matrix. When $p$ is odd, we have a knot, and when it is even, we
have a link.  In the $SU(N)_{q}$ case, these invariants are
proportional to HOMFLY polynomials depending on the variables
$q = e^{\pi i/(k+N)}$ and $A=q^{N}$ \cite{Zodinmawia:2011oya,
  Nawata:2013qpa,Gu2014b}. The proportionality factor depends on the
choice of framing for the Wilson loops, but are canonically chosen to
not depend explicitly on $N$, except through $A$ \cite{Gu2014b}. When
$N=2$, we get Jones polynomials as a special case of the HOMFLY ones.

In this paper, we look for the appropriate $S$ and $T$ matrices for
fields labeled by representations of $W_{N}$ algebra.  Explicit
calculation shows that these matrices are not properly normalized to
give the usual quantum Racah matrices
\cite{Ponsot1999,Teschner2014,Gu2014b}. In order to have an explicit
representation of the braid group, we have to find a conformal block
normalization such that $S$ is an unitary hermitian matrix, that is,
\begin{equation}
  \label{eq:unitaryconstraint}
  SS^{\dagger} = \mathbbold{1}, \quad S = S^{\dagger},\entao S^{2}=\mathbbold{1}.
\end{equation} 
This property will allow us to fix the normalization. The $S$ matrix
is not hermitian in general, but it will be valid in our particular
two-dimensional case. For knots, we have two types of crossing and
braiding matrices corresponding to the parallel case $(S, T)$ and to
the antiparallel case $(\bar{S}, \bar{T})$.  We also need that the
Yang-Baxter equation be satisfied for certain 3-strand moves.  These
matrices must obey the unknot constraint \cite{Mironov2015}
\begin{equation}
\label{eq:knotconstraint}  
S\bar{T}\bar{S} = T^{-1}S\bar{T}^{-1},
\end{equation}
and the Yang-Baxter equation
\begin{equation}
  \label{eq:yang-baxter}
  S\bar{T}\bar{S}\bar{T}STS=TS\bar{T}\bar{S}\bar{T}.
\end{equation}
These two equations will then allow us to choose a correct framing for
the $T$ matrices below.

\subsection{Normalization of Conformal Blocks}
\label{sec:norm-conf-blocks}

In general, a four-point correlation function of primary fields in a
CFT with symmetry algebra $g$ can be written as
\begin{equation}
\label{eq:15}
  \left\langle V_{\alpha_{1}}(z)V_{\alpha_{2}}(1)V_{\alpha_3}(0)V_{\alpha_4}(\infty)\right\rangle =
  (\mathcal{G}^{s})^{\dagger}\mathcal{M}^{s}\mathcal{G}^{s} = (\mathcal{G}^{t}) ^{\dagger}\mathcal{M}^{t}\mathcal{G}^{t},
\end{equation}
with $\alpha_{i}$ being $g$-valued vectors labelling the primaries,
$\mathcal{G}^{k}=\mathcal{G}^{k}(z)$ are the conformal blocks in the
$k=s,t$-channels and $\mathcal{M}^{k}$ are constant matrices (more
generally, bilinear transformations) formed by the product of
structure constants of each channel. The space of conformal blocks
$\mathcal{H}^{k}$ is finite dimensional depending if one or more
fields in \eqref{eq:15} are degenerate. This is always the case for
rational conformal field theories. In the following, we suppose that
the matrices $\mathcal{M}^{k}$ are diagonal, which is not necessarily
true in general because of non-trivial monodromy properties of
higher-spin conformal blocks \cite{Chang2012}. We shall review the
calculation of $W_{N}$ correlators in the next section.

We want to change the normalization of the conformal blocks in such a
way that the S matrix is an unitary hermitean matrix.  First, let us
define the new blocks as $\mathcal{F}^{k} = N_{k}\mathcal{G}^{k}$
(with no index summation), where $k=s,t$ denotes the respective
channels and $N_{k}$ are diagonal normalization matrices. Also, we
have that $\mathcal{G}^{s} = S\mathcal{G}^{t}$ and, using this in
\eqref{eq:15}, we get
\begin{equation}
\label{eq:16}
  S^{\dagger}\mathcal{M}^{s}S = \mathcal{M}^{t},
\end{equation}
Changing the S matrix to $\tilde{S}$ in the new normalization, we
write $S = N_{s}^{-1}\tilde{S}N_{t}^{\phantom{1}} $. If we plug this into
\eqref{eq:16} we obtain
\begin{equation}
\label{eq:21}
  \mathcal{M}^{k} = \alpha N^{\dagger}_{k}N^{\phantom{\dagger}}_{k},\quad \alpha \in \mathbb{R}.
\end{equation}
In the following,  we set $\alpha=1$ as it depends only on the overall
normalization of the correlation function. 

Now, as will be clear below, we suppose that $\mathcal{H}^{k}$ is
two-dimensional. Then we can parametrize the normalization matrices as
\begin{equation}
\label{eq:parametrization}
  N_{k} = \delta_{k}
  \begin{pmatrix}
    \zeta_{k}^{-1} & 0\\
    0 & \zeta_{k}
  \end{pmatrix}.
\end{equation}
Setting $\mathcal{M}^{k} = \text{diag}(C_{1}^{k},C_{2}^{k})$, we get
\begin{equation}
\label{eq:17}
  \frac{C_{1}^{k}}{C_{2}^{k}} = \frac{1}{|\zeta_{k}|^{4}},
\end{equation}
where $C_{j}^{k}$ corresponds to the products of structure constants
in the $k$-channel appearing in \eqref{eq:15}.  We determine
$\delta_{k}$ up to a phase by
\begin{equation}
\label{eq:18}
  \Tr \mathcal{M}_{k} = C_{1}^{k}+C_{2}^{k} =
                      |\delta_{k}|^{2}(|\zeta_{k}|^{-2}+|\zeta_{k}|^{2}).
\end{equation}
From equations \eqref{eq:17} and \eqref{eq:18} we thus get
\begin{equation}
  |\zeta_{k}| = 
  \left(\frac{C_{2}^{k}}{C_{1}^{k}}
  \right)^{1/4},\quad |\delta_{k}| =
  \left(C_{1}^{k}C_{2}^{k}\right)^{1/4}.
\end{equation}
Therefore, if we can find the products of structure constants
$C_{j}^{k}$, we can fix the normalization. When we know the $S$
matrix, we can go the other way around and use it to find the
structure constants, which is at the core of the bootstrap
approach. 

The discussion presented here should be compared with
\cite{Teschner2014}, where its authors define the proper normalization
of conformal blocks to obtain the Racah-Wigner b-$6j$ symbols
associated to the modular double of
$\mathcal{U}_{q}(\mathrm{sl}(2,\mathbb{R}))$ with $q= e^{i\pi b^{2}}$.
The representations of this quantum group will explicitly appear below
in our approach of loop invariants in the Virasoro case.

\subsection{Knot and Link Invariants from Virasoro Representations}
\label{ssec:virasoro-case}

Here we briefly review how to calculate Jones polynomials from S and T
matrices related to fields in Virasoro representations. We start with
a few definitions following \cite{Ribault:2014aa}, for example. Chiral
vertex operators $V_{\alpha}$ are labeled by charge vectors
$\alpha\equiv \alpha_{r,s}$, which also label the conformal dimension
of $V_{\alpha}$
\begin{equation}
  \Delta_{(r,s)} = \alpha_{r,s}(Q-\alpha_{r,s}),\quad \alpha_{r,s} =
  \frac{1}{2}\{(1-r)b + (1-s)b^{-1} \},
\end{equation}
where $Q= b + 1/b$. The integers $r,s$ label unitary irreducible Verma
modules with central charge $c$ less than one, appearing in the
minimal models. Let us calculate the conformal blocks of the 4-point
correlation function holomorphic part
\begin{equation}
  F_{\hat\alpha}(z,z_{1},z_{2},z_{3}) \equiv\langle
  V_{\alpha}(z)V_{\alpha_{1}}(z_{1})V_{\alpha_{2}}(z_{2})V_{\alpha_{3}}(z_{3})\rangle,
\end{equation}
where $\hat\alpha = (\alpha,\alpha_{1},\alpha_{2},\alpha_{3})$ and the
field $V_{\alpha}$ is degenerate at level 2, i.e., $(r,s) = (1,2)$ or
$(2,1)$. The charge vectors must obey the neutrality condition
\begin{equation}
  \alpha + \sum_{i=1}^{3}\alpha_{i} + mb+nb^{-1} = Q,\quad m,n\in \mathbb{Z}_{+}.
\end{equation}
One of the fields being degenerate at level 2 implies either one of
the null vector conditions below
\begin{align}
(  L_{-2} + b^{2} L_{-1}^{2})V_{\alpha} &= 0\quad\text{for}\quad (r,s) 
= (1,2),\\
(  L_{-2} + \tfrac{1}{b^{2}}L_{-1}^{2})V_{\alpha} &= 0\quad\text{for}\quad (r,s) 
= (2,1)\label{eq:19} . 
\end{align}
Let us focus on the choice $(r,s) = (2,1)$, such that
$\alpha = -\tfrac{b}{2}$. Using the conformal Ward identity, we find
the action of the Virasoro operators on the correlators,
which then implies in the BPZ equation
\begin{equation}
  \left(
    \frac{1}{b^{2}}\Dpfrac[2]{}{z} + \sum_{i=1}^{3}
\left[
  \frac{1}{z-z_{i}}\Dpfrac{}{z_{i}}+\frac{\Delta_{i}}{(z-z_{i})^{2}}
\right]
  \right) F_{\hat\alpha}(z,z_{1},z_{2},z_{3}) =0.
\end{equation}
Using $SL(2,\mathbb{C})$ invariance, we set
$(z_{1},z_{2},z_{3})\rightarrow (0,\infty,1)$ and then get
\begin{equation}
  \frac{1}{b^{2}}F''_{\hat\alpha}(z) +
  \frac{2z-1}{z(1-z)}F'_{\hat\alpha}(z)+
  \left(\frac{\Delta_{1}}{z^{2}}+\frac{\Delta_{3}}{(1-z)^{2}}
  +\frac{\Delta_{(2,1)}+\Delta_{1}-\Delta_{2}+\Delta_{3}}{z(1-z)}\right) F_{\hat\alpha}(z) =0.
\end{equation}
Finally, we put this equation into hypergeometric form
by making
$F(z) = z^{b\alpha_{1}}(1-z)^{b\alpha_{3}}G(z)$
\begin{equation}
\label{eq:7}
\{  z(1-z)\Dpfrac[2]{}{z}+[C -(A+B+1)z]\Dpfrac{}{z}-AB\}G(z) =0,
\end{equation}
where
\begin{align}
\label{eq:11}
  A &= \frac{1}{2}+b(\alpha_{1}+\alpha_{3}-Q)+
  b(\alpha_{2}-\frac{Q}{2}),\\[5pt]
  B &= \frac{1}{2}+b(\alpha_{1}+\alpha_{3}-Q)-
  b(\alpha_{2}-\frac{Q}{2}),\\[5pt]
  C &= 1+b(2\alpha_{1}-Q).
\end{align}
The solutions of \eqref{eq:7} are given by hypergeometric functions
$\sideset{_2}{_{1}}\genhyp(A,B;C|z)$. If we label the conformal blocks
as $\mathcal{F}^{k}=(F^{k}_{1}\;\; F^{k}_{2} )$, where $k =s,t$
denotes the channels, we have the s-channel conformal blocks
\begin{subequations}
\label{eq:13}
  \begin{align}  
    F^{s}_{1}(z) &=
                   z^{b\alpha_{1}}(1-z)^{b\alpha_{3}}\sideset{_2}{_{1}}\genhyp(A,B;C|z),\label{eq:13a}\\
    F^{s}_{2}(z) &=
                   z^{b(Q-\alpha_{1})}(1-z)^{b\alpha_{3}}\sideset{_2}{_{1}}\genhyp(A-C+1,B-C+1;2-C|z),
\label{eq:13b}
  \end{align}
\end{subequations}
and the t-channel conformal blocks
\begin{subequations}
  \begin{align}
    F^{t}_{1}(z) &=
                   z^{b\alpha_{1}}(1-z)^{b\alpha_{3}}\sideset{_2}{_{1}}\genhyp(A,B;A+B-C+1|\;
                   1-z),\\
    F^{t}_{2}(z) &=
                   z^{b\alpha_{1}}(1-z)^{b(Q-\alpha_{3})}\sideset{_2}{_{1}}\genhyp(C-A,C-B;C-A-B+1|\;
                   1-z).
  \end{align}
\end{subequations}
The crossing matrix $S$ is given by
$\mathcal{F}^{s} = S\mathcal{F}^{t}$ where
\begin{equation}
\label{eq:connectionformula}
  S=\begin{pmatrix}
    \frac{\Gamma(C)\Gamma(C-A-B)}{\Gamma(C-A)\Gamma(C-B)} &
    \frac{\Gamma(C)\Gamma(A+B-C)}{\Gamma(A)\Gamma(B)} \\[5pt]
    \frac{\Gamma(2-C)\Gamma(C-A-B)}{\Gamma(1-A)\Gamma(1-B)} & \frac{\Gamma(2-C)\Gamma(A+B-C)}{\Gamma(A-C+1)\Gamma(B-C+1)} 
  \end{pmatrix}.
\end{equation}
In the particular case of the other fields also in the fundamental
representation, $\alpha_{1}=\alpha_{3} = -b/2$ and
$\alpha_{2}=Q-\alpha_{1}$. Thus, we get
\begin{align}
\label{eq:5}
 S= \left(
\begin{array}{cc}
 \frac{1}{[2]}& \frac{\Gamma \left(-2 b^2\right) \Gamma \left(-2 b^2-1\right)}{\Gamma \left(-3 b^2-1\right) \Gamma \left(-b^2\right)}\\[5pt]
 \frac{\Gamma \left(2 b^2+1\right) \Gamma \left(2 b^2+2\right)}{\Gamma
  \left(b^2+1\right) \Gamma \left(3 b^2+2\right)} & -\frac{1}{[2]} \\
\end{array}
  \right)
=
  \begin{pmatrix}
    \frac{1}{[2]} &  \frac{\sqrt{[3]}}{[2]}\zeta^{2}\\[5pt]
    \frac{\sqrt{[3]}}{[2]}\zeta^{-2}& -\frac{1}{[2]}
  \end{pmatrix}
=
  U
  \begin{pmatrix}
  \frac{1}{[2]} & \frac{\sqrt{[3]}}{[2]}\\[5pt]
  \frac{\sqrt{[3]}}{[2]}& -\frac{1}{[2]}
  \end{pmatrix}
   U^{-1}
\end{align}
where $q=e^{i\pi b^{2}}$, $[N]=\frac{q^{N}-q^{-N}}{q-q^{-1}}$ and
\begin{equation}
  U=
  \begin{pmatrix}
    \zeta & 0\\
    0 & \zeta^{-1}
  \end{pmatrix},\quad \zeta^{2} =\frac{\sqrt{[3]}}{[2]} \frac{ \Gamma
    \left(b^2+1\right) \Gamma
    \left(3b^2+2\right)}{\Gamma \left(2 b^2+1\right) \Gamma
    \left(2b^2+2\right)}.
\end{equation}
It is now easy to check that $S^{2}=\mathbbold{1}$, as required by
\eqref{eq:unitaryconstraint}.  The $s$-channel braiding matrix $T$ is
obtained by making a half turn around zero in \eqref{eq:13}
\begin{equation}
  T = N(q)
  \begin{pmatrix}
    q^{-1/2} &0\\
    0& -q^{3/2}
  \end{pmatrix},
\end{equation}
where $N(q)$ is an overall normalization.  The Yang-Baxter equation in
this case is $(ST)^{3} =\mathbbold{1}$ and this let us choose
$N(q) = -q^{-1/2}$. 
Summing up, the matrices representing the braid group $B_{3}$ colored
with $SU(2)$ representations are
\begin{equation}
 S=  \begin{pmatrix}
  \frac{1}{[2]} & \frac{\sqrt{[3]}}{[2]}\\[5pt]
  \frac{\sqrt{[3]}}{[2]}& -\frac{1}{[2]}
  \end{pmatrix},\quad T = 
  \begin{pmatrix}
    -\frac{1}{q} &0\\
    0& q
  \end{pmatrix}.
\end{equation}
We can construct knot and link polynomials by starting with the
$t$-channel conformal blocks, braiding representations in the
$s$-channel $k$ number of times and then going back to the
$t$-channel, like in fig.~\ref{fig:innerprod}. In this way, we get
\begin{equation}
\label{eq:20}
  S T^{p} S = \frac{1}{[2]^{2}}\left(
\begin{array}{cc}
 q^{ p} [3]+(-1)^{p}q^{- p} & -\left(q^{ p}+(-1)^{p+1}q^{- p}\right) \sqrt{[3]} \\
-\left(q^{ p}+(-1)^{p+1}q^{- p}\right) \sqrt{[3]} &  q^{ p} +(-1)^{p}q^{- p}[3] \\
\end{array}
\right).
\end{equation}
According to the fusion rules, the conformal block with zero
intermediate weight is \eqref{eq:13b} and, thus, the knot/link
invariant of interest is the second diagonal component of
\eqref{eq:20}, which is an \emph{unreduced} Jones invariant. The
reduced Jones knot polynomial is defined as the ratio of the unreduced
polynomial and the \emph{unknot} in the same representation
\begin{equation}
  J^{C,2k+1}_{R}(q) = \frac{Z^{t}_{R}(2k+1|C)}{Z^{t}_{R}(1|C)}=
  \frac{(ST^{2k+1}S)_{22}}{(STS)_{22}}= -q^{4}+q^{2}+q^{-2}.
\end{equation}
Notice that the second diagonal component can be recovered from the
first by making $q \rightarrow -1/q$. A more extensive discussion of
the types of knots and link invariants calculated in this way is given
in \cite{D.Galakhov:2015aa,Mironov2015}.

\section{Toda Field Theory and $W_{N}$ Conformal Blocks}
\label{sec:toda-field-theory}

In this section, we review the machinery of Toda field theory,
discussed in \cite{Fateev2007a}, in order to generalize the
construction above for $S$ and $T$ matrices in $W_{N}$
models. Specifically, we fix two fields in the four-point function to
be in the fundamental representation of $SU(N)$ paired with its
conjugate representation and deduce some results for the other fields
in an more general representation. This section follows mostly the
definitions and conventions of \cite{Fateev2007a}. Other relevant
references about correlators in $W_{N}$ models are
\cite{V.A.Fateev:2008aa,Chang2012}.

The generalization of Liouville theory extending Virasoro to $W_{N}$
algebra is called Toda field theory (TFT). The basic field is a scalar
field $\varphi = \sum_{i=1}^{N-1}\varphi_{i}e_{i}$, where $e_{i}, i=1,...,N-1$, are
the simple roots of  $su(N)$ algebra. The most important information
about the algebra is contained in the \emph{Cartan matrix}, $K_{ij}$,
defined by the inner product of the simple roots,
$K_{ij}=(e_{i},e_{j})$. From the inner product, one can define the
dual weight space in terms the \emph{fundamental weights} $\omega_{k}$
by $(\omega_{k},e_{j}) = \delta_{kj}$ and the \emph{quadratic form} of
the algebra by $(\omega_{i},\omega_{j}) = K_{ij}^{-1}$. A highest
weight $\lambda$ takes the form
\begin{equation}
  \label{eq:2}
  \lambda = \sum_{i=1}^{N-1}\lambda_{i}\omega_{i},
\end{equation}
where
$ (\lambda_{1},\hdots,\lambda_{N-1}) \in \mathbb{Z}^{N-1}_{\geq 0}$
are called Dynkin labels. In particular, the conjugate representation
of $\lambda$ is represented by
$\bar{\lambda} = \sum_{i=1}^{N-1}\lambda_{N-i}\omega_{i}$.

To each highest weight we can associate a \emph{partition}
$\lambda = \{\ell_1;\ell_2;...;\ell_{N-1}\}$ where
$\ell_i = \lambda_{i}+\lambda_{i+1}+...+\lambda_{N-1}$. We then
associate a \emph{Young tableau} to the partition by assigning
$\ell_i$ boxes to the $i$-th row of the tableau. Some simple examples
are
\begin{align*}
  F_{1}&=(1,0,0,...,0)\;\sim\;\yng(1)
  \\[5pt]
 F_{2}&= (2,0,0,...,0)\;\sim\;\yng(2)
  \\[5pt]
A_{2}&= (0,1,0,...,0)\;\sim\;\yng(1,1)
\end{align*}
where the first example is the fundamental representation, the second
a symmetric representation and the third an antisymmetric
representation. Young diagrams are useful to build up tensor product
representations and thus analyze possible states of fusion rules. For
more details, see \cite{DiFrancesco1997}, for example. To find all the
states in an irreducible module with highest weight $\lambda$, we
subtract all possible combinations of simple roots $e_{i}$ up to
$\lambda_{i}e_{i}$ for each positive $\lambda_{i}$. Then we repeat the
process with the new weights until there is no way to produce a new
weight with positive Dynkin label. In the case of the fundamental
representation, the weights are expressed as
\begin{equation}
  h_{k} = \omega_{1}-\sum_{i=1}^{k-1}e_{i},\quad k=1,...,N.
\end{equation}

The TFT action on a Riemann surface with reference metric
$\hat{g}_{ab}$ and scalar curvature $\hat{R}$ is given by
\begin{equation}
  S_{TFT} = \int 
  \left(
    \frac{1}{8\pi}\hat{g}^{ab}(\del[a]\varphi,\del[b]\varphi)+\frac{(Q,\varphi)}{4\pi}\hat{R}
    + \mu \sum_{k=1}^{N-1}e^{b(e_{k},\varphi)}
  \right)\sqrt{\hat{g}}\;d^{2}x,
\end{equation}
where $\mu$ is the cosmological constant and $ Q $ is the
background charge. To ensure conformal invariance, we must set the
charge to be
\begin{equation}
 Q= (b+1/b)\rho,\quad \rho = \sum_{k=1}^{N-1}\omega_{k},
\end{equation}
where $\rho$ is the \emph{Weyl vector} of the algebra. The theory is
invariant under symmetries generated by the currents $W^{k}(z)$ with
spins $k=2,3,4,...,N$ and its antiholomorphic counteparts. The current
$W^{2}(z)\equiv T(z)$ is equal to the energy-momentum tensor and the
other are higher-spin currents. Together, these currents generate the
$W_{N}$ algebra containing the Virasoro algebra with central charge
$c=N-1+12Q^{2}$.

 

The Toda correlators can be calculated in the Coulomb gas formalism by
introducing the chiral vertex operators
$V_{\alpha} = e^{(\alpha,\varphi)}$. The OPE with the currents
$W^{k}(z)$ classify the states in terms of the quantum numbers
$w^{k}(\alpha)$, $k=2,3,...,N-1$. In particular, the conformal weight
is given by
\begin{equation}
  w^{(2)}(\alpha) = \Delta(\alpha) = \frac{(\alpha,2Q-\alpha)}{2}.
\end{equation}
The $w^{k}(\alpha)$ are invariant under the $su(N)$ Weyl group. After
a Weyl reflection, the field $V_{\alpha}$ acquires a reflection
amplitude \cite{Fateev2007a}. As an example, the conjugate
representation $\bar\alpha$ is equivalent to $2Q-\alpha$ under the
longest Weyl reflection. This changes the correlation function by a
multiplicative factor which will not be relevant to calculate reduced
polynomials, as overall factors cancel.

All of this corresponds to the general Toda theory. The $W_N$ minimal
model can be realized as the coset model
$SU(N)_k\oplus SU(N)_1 / SU(N)_{k+1}$.  After imposing the constraint
in the root lattice, we find that the primaries are labeled by
$(\rho;\nu)\equiv (\Lambda;\tilde{\Lambda})$ (level $k$ and $k+1$
respectively). The details of this construction can be found in
\cite{Bouwknegt1993a,Chang2012}, for example. As we are going to see
below, the $W_{N}$ conformal blocks can be obtained by taking the
residue of Toda conformal blocks, similarly to the Virasoro and
Liouville case.


The Toda 3-point function with one semi-degenerate field was first
calculated in \cite{Fateev:2005aa} and a general formula using AGT
relation was proposed in \cite{Mitev:2014aa,Isachenkov:2014aa}. In the
Virasoro case, knowledge of the two and three-point functions allow us
to obtain multipoint correlators by the conformal bootstrap
\cite{Belavin1984}. As we saw in section~\ref{ssec:virasoro-case}, the
four-point function is completely determined by setting one of the
fields to be completely degenerate. However, for the $W_{N}$ case this
is not enough \cite{Bowcock1994a,Fateev:2005aa}. The structure of the
Verma modules is more constrained by the extra higher-spin symmetries
and we need to fix another field to be in a semi-degenerate state
$\alpha = \kappa\omega_{N-1}$, where $\kappa$ is an arbitrary constant
\cite{Fateev2007a}.  Here we shall restrict our discussion to this
semi-degenerate case. For more details, see
\cite{Fateev:2005aa,Fateev2007a,V.A.Fateev:2008aa}.

Three-point correlators are constrained by conformal invariance to be
\begin{equation}
   \langle \vertex{1}{\alpha_{1}}\vertex{2}{\alpha_{2}}\vertex{3}{\alpha_{3}}
\rangle = \frac{C(\alpha_{1},\alpha_{2},\alpha_{3})}{|z_{12}|^{2(\Delta_{1}+\Delta_{2}-\Delta_{3})}|z_{13}|^{2(\Delta_{1}+\Delta_{3}-\Delta_{2})}|z_{23}|^{2(\Delta_{2}+\Delta_{3}-\Delta_{1})}}.
\end{equation}
Analysis of the Coulomb integral for calculating this function shows that
the structure constants $C(\alpha_{1},\alpha_{2},\alpha_{3})$ have
poles when the screening condition is satisfied
\begin{equation}
(2Q-  \sum_{i=1}^{3}\alpha_{i},\omega_{k} )=
bs_{k}+b^{-1}\tilde{s}_{k},\quad s_{k},\tilde{s}_{k} \in
\mathbb{Z}_{\geq 0}.
\end{equation}
Taking the residues of the correlator in those poles gives the $W_{N}$
structure constants \cite{Fateev2007a,Chang2012}.  Those can be
expressed in terms of complicated Coulomb integrals, but in some
simple cases, like when one of the fields is semi-degenerate, they can
be written in terms of known special functions
\cite{Fateev2007a}. Defining
\begin{equation}
 C^{\alpha_{3}}_{\alpha_{1},\alpha_{2}} \equiv   C(\alpha_{1},\alpha_{2}, 2Q-\alpha_{3}),
\end{equation}
our  particular case of interest is when one of the fields above is in
the fundamental representation
\begin{equation}
\label{eq:structureconstants}
  C^{\alpha_{1}-bh_{k}}_{-b\omega_{1},\alpha_{1}} =
  \left (-\frac{\pi \mu}{\gamma(-b^{2})}\right)^{k-1}\,\prod_{i=1}^{k-1}\frac{\gamma(b(\alpha_{1}-Q,h_{i}-h_{k}))}{\gamma(1+b^{2}+b(\alpha_{1}-Q,h_{i}-h_{k}))},
\end{equation}
where $\gamma(z) = \Gamma(z)/\Gamma(1-z)$. With this formula, we can
show that the fusion rule of $V_{-b\omega_{1}}$ and
$V_{\kappa \omega_{N-1}}$ has only two fields. In particular, we have
only two intermediate states in a channel with a fundamental and
anti-fundamental field and thus the space of conformal blocks is
two-dimensional.

Now let us consider the four-point correlator after fixing three
points by global $\mathrm{SL}(2,\mathbb{C})$ invariance
\begin{equation}
  \label{eq:4pointcorrelator}
  \langle V_{-b\omega_{1}}(z,\bar
  z)V_{\alpha_{1}}(0)V_{\alpha_{2}}(\infty)V_{-b\omega_{N-1}}(1) \rangle
  = |z|^{2b(\alpha_{1},\omega_{1})}|1-z|^{-\frac{2b^{2}}{N}}G(z,\bar z),
\end{equation}
where $\alpha_{2}=2Q-\alpha_{1}$ and, in the $s$-channel expansion,
\begin{equation}
  \label{eq:6}
  G(z,\bar z) = \sum_{j=1}^{N}
  C^{\alpha_{1}-bh_{j}}_{-b\omega_{1},\alpha_{1}}C(\alpha_{1}-bh_{j},\alpha_{2},-b\omega_{N-1})G_{j}(z)G_{j}(\bar
  z).
\end{equation}
The summation over intermediate states follows from the fusion rules
\cite{Fateev1988,Fateev2007a}
\begin{equation}
  V_{-b\omega_{k}}V_{\alpha}=\sum_{s}C^{\alpha-bh_{s}^{(k)}}_{-b\omega_{k},\alpha}
  \left[
    V_{\alpha-bh_{s}^{(k)}}
  \right],
\end{equation}
where $h_{s}^{(k)}$ are the weights of the representation with highest
weight $\omega_{k}$.  

The conformal blocks $G_{j}(z)$ satisfy the generalized hypergeometric
equation
\begin{equation}\label{hg-eq}
	\left[z\prod_{k=1}^N\left(\theta+A_k\right)-\theta\prod_{k=1}^{N-1}\left(\theta+B_k-1\right)\right]G_{j}(z)=0,
\end{equation}
where $\theta = z\frac{d}{dz}$ and the coefficients $A_k$ and $B_k$
are given by
\begin{equation}
\label{eq:hypergconstants}
	\begin{gathered}
	A_k=-b^2+b(\alpha_1-Q,e_1+\cdots+e_{k-1}),\\
	B_k=1+b(\alpha_1-Q,e_1+\cdots+e_k).
	\end{gathered}
\end{equation}
In terms of the generalized hypergeometric function,
\begin{align}
G_{1} &= \sideset{_{N}}{_{N-1}}\genhyp
  \left(
    \begin{gathered}
    \scriptstyle A_{1}\,...\, A_{N}\\[-8pt]
    \scriptstyle  B_{1}\,...\, B_{N-1}
    \end{gathered}
  \bigg|\,z \right),\\[5pt]
  G_{j}&=z^{1-B_{j}}\sideset{_{N}}{_{N-1}}\genhyp
  \left(
    \begin{gathered}
      \scriptstyle 1-B_{j}+A_{1}\,...\, 1-B_{j}+A_{N}\\[-7pt]
      \scriptstyle 1-B_{j}+B_{1}\,...\, 2-B_{j}\,...\, 1-B_{j}+B_{N-1}
    \end{gathered}
  \bigg|\,z \right),\quad 1< j \leq N.
\end{align}
By consistency between $s$ and $u$-channel expansions, we find
\begin{equation}
\label{eq:connectionformula}
\frac{
  C^{\alpha_{1}-bh_{1}}_{-b\omega_{1},\alpha_{1}}C(\alpha_{1}-bh_{1},\alpha_{2},-b\omega_{N-1})}{
  C^{\alpha_{1}-bh_{k}}_{-b\omega_{1},\alpha_{1}}C(\alpha_{1}-bh_{k},\alpha_{2},-b\omega_{N-1})}
=
\frac{\prod_{j=1}^{N}\gamma(A_{j})\gamma(B_{k-1}-A_{j})}{\prod_{j=1}^{N-1}\gamma(B_{j})}\frac{\prod_{j\neq
  k-1}\gamma(1+B_{j}-B_{k-1})}{\gamma(B_{k-1}-1)},
\end{equation}
which follows from connection formulas of generalized hypergeometric
functions \cite{Noerlund1955,Fateev2007a}. Those are the basic
equations used to find three-point functions with one partially
degenerate field.

\section{Crossing and Braiding matrices in $W_N$ models}
\label{sec:knot-invariants}
\newcommand{\nt}{\tilde{n}} Now that we know the $W_{N}$ conformal
blocks, we can analyze particular cases to calculate crossing $S$ and
braiding matrices $T$ with one fundamental and one anti-fundamental
field, as in \eqref{eq:4pointcorrelator}. Let us start by taking a
highest-weight state in the form
\begin{equation}
\alpha_{1}=-b\Lambda-b^{-1}\tilde{\Lambda},
\end{equation}
related to the pair of representations $(\Lambda;\tilde{\Lambda})$
labelling a $W_{N}$ primary, where
\begin{equation}
  \Lambda = \sum_{i=1}^{N-1}n_{i}\omega_{i},\quad \tilde{\Lambda} =
  \sum_{i=1}^{N-1}\nt_{i}\omega_{i},\quad n_{i},\tilde{n}_{i}\in
  \mathbb{Z}_{\geq 0}.
\end{equation}
We are going to see that if a number
$r$ of labels $n_{i}$ or $\nt_{i}$ are different from zero, then
\eqref{hg-eq} is reducible to a lower order hypergeometric
equation. First, notice that
\begin{equation}
  B_{k} = A_{k+1}+b^{2}+1 = A_{k} +b(\alpha_{1},e_{k}),\quad k=1,...,N-1 .
\end{equation}
As $z(\theta+A)f(z) = (\theta+A-1)zf(z)$, it is easy to see that for
each $B_{k}=A_{k}$ we can factor out a term $(\theta+B_{k}-1)$ from
eq.~\eqref{hg-eq}, effectively reducing its order. For the particular
$\alpha_{1}$ we are considering,
\begin{equation}
  B_{k} = A_{k}-n_{k}b^{2}-
  \nt_{k}.
\end{equation}
Therefore, $B_{k} = A_{k}$ except for $n_{k},\nt_k\neq 0$. If $r$
labels $n_{i}$ or $\nt_{i}$ are different from zero, the $(N,N-1)$
generalized hypergeometric operator factorizes to a product
$D_{N,N-1} = P_{N-r-1}D_{r+1,r}$ of an order $N-r-1$ operator and a
$(r+1,r)$ hypergeometric operator. This proves our assertion. Finally,
we can explicitly write the $A_{k}$ as
\begin{equation}
\label{eq:9}
  A_{k} = -(\ell_{1} - \ell_k +k)b^{2} -(\tilde{\ell}_{1}-\tilde{\ell}_{k}+k-1),
\end{equation}
where $\ell_{k},\tilde{\ell}_{k}$ are the number of boxes in the
$k$-th row of the representation $\Lambda,\tilde{\Lambda}$,
respectively.

The connection matrices between $z=0$ and $z=1$ for higher-order
hypergeometric equations are not easy to find and we will not consider
those in this paper. However, for a reduction to Gauss hypergeometric
equation, we have explicit formulas like
\eqref{eq:connectionformula}. This correspond to the case of
\emph{rectangular} representations
$\alpha_1=- (nb+\tilde{n}b^{-1}) \omega_m$. Here $m$ corresponds to
the number of rows and $n,\tilde{n}$ the number of columns of the
Young diagram of $\Lambda, \tilde{\Lambda}$ respectively. The field
$\alpha_{2}=2Q-\alpha_{1}$ is Weyl equivalent to
$\bar{\alpha}_{1}=-(nb+\tilde{n}b^{-1})\omega_{N-m}$ and then the
correlation function differs from \eqref{eq:4pointcorrelator} by an
overall reflection amplitude, which will not be relevant for
us. 
In this case, eq. \eqref{eq:9} becomes
\begin{align}
	A_{k} &= -( n H(k-m-1) +k)b^{2}-(\nt H(k-m-1)+k-1),
\end{align}
where $H(k)$ is the step function. We have that
$B_{k} = A_{k} - (nb^{2}+\nt)\delta_{k}^{m}$ for $1\leq k < N$,
therefore $B_{k} = A_{k}$ for all $ k\neq m$ and \eqref{hg-eq} reduces
to a second order hypergeometric equation \eqref{eq:7} with parameters
\begin{gather}
	A=A_m=-mb^{2}-(m-1),\quad B=A_N=-(N+n)b^2-(N+\nt-1),\\
	C=B_m=-(m+n)b^2-(m+\nt-1).
\end{gather}
The $S$ matrix now is
\begin{equation}
  S=
\begin{pmatrix}
  \frac{\Gamma \left(-(n+m) b^2- (m+\nt-1)\right) \Gamma \left(N
      b^2+N-1\right)}{\Gamma \left(-n b^2-\nt \right) \Gamma
    \left((N-m)\left(b^2+1\right) \right)} & \frac{\Gamma \left(-(n+m)
      b^2- (m+\nt-1)\right) \Gamma \left(-Nb^2- (N-1)\right)}{\Gamma
    \left(-mb^2-(m-1)\right) \Gamma \left(-(N+n) b^2-(N+\nt-1)\right)} \\[10pt]
  \frac{\Gamma \left((m +n) b^2+m+\nt+1\right) \Gamma \left(N
      b^2+N-1\right)}{\Gamma \left(m (b^2+1)\right) \Gamma \left((N+n)
      b^2+N+\nt \right)} & \frac{\Gamma \left((m+n) b^2+m+\nt+1\right) \Gamma \left(-Nb^2- (N-1)\right)}{\Gamma \left(n b^2+\nt+1\right) \Gamma \left(-(N-m) b^2-(N-m-1)\right)}
\end{pmatrix}
\end{equation}
and, as $S = N_{s}^{-1}S_{m,n}N_{t}$, we have that
\begin{align}
\label{eq:8}
  S=\delta_{m,n}\begin{pmatrix}
    \xi_{m,n}^{2} s_{1} & \zeta_{m,n}^{2}s_{2} \\[5pt]
    \zeta_{m,n}^{-2}s_{2}& -\xi_{m,n}^{-2} s_{1}
	\end{pmatrix}  = \delta_{m,n}\,U R\, S_{m,n}\, R U^{-1}, 
\end{align}
where
\begin{gather}
R =
\begin{pmatrix}
  \xi_{m,n} & 0\\
  0 & \xi_{m,n}^{-1}
\end{pmatrix},\quad 
U =
\begin{pmatrix}
 \zeta_{m,n} & 0\\
 0 & \zeta_{m,n}^{-1}
\end{pmatrix}.
\end{gather}
Relating with the parametrization \eqref{eq:parametrization}, we have
that $\delta_{m,n}=\delta_{t}/\delta_{s}$ and
\begin{align}
  N_{t} &= \delta_{t}RU^{-1} =
  \delta_{t}\begin{pmatrix}
    \gamma^{-1}_{t} &0\\
    0& \gamma_{t}
  \end{pmatrix},\quad \gamma_{t} = \frac{\zeta_{m,n}}{\xi_{m,n}},\\[10pt]
N_{s} &= \delta_{s}R^{-1}U^{-1}=
    \delta_{s}  \begin{pmatrix}
      \gamma^{-1}_{s} & 0\\
      0 & \gamma_{s}
  \end{pmatrix}, \quad \gamma_{s} = \zeta_{m,n}\,\xi_{m,n},
\end{align}
fixing the normalization in the $t$ and $s$-channel conformal blocks,
now defined as $\mathcal{F}^{k} = N_{k}\mathcal{G}^{k}$. Choosing
$\det S_{m,n} = -1$, the parametrization coefficients are given by
\begin{align}
  \delta_{m,n}^{2} &= -\det S = \frac{(m+n)b^{2}+m+\nt}{Nb^{2}+N-1},\\[5pt]
  \zeta_{m,n}^{2} &= \left(\frac{S_{12}}{S_{21}} \right)^{\frac{1}{2}}\nonumber
  \\[5pt]
                   &=\delta^{-1}_{m,n}
                     \sqrt{\frac{[N+n][m]}{[N][n+m]}}\frac{\Gamma(m(b^{2}+1))\Gamma((N+n)b^{2}+N+\nt)}{\Gamma((N(b^{2}+1)))\Gamma((m+n)b^{2}+m+\nt)},\\[5pt] 
  \xi_{m,n}^{2} &=  \left(-\frac{S_{11}}{S_{22}}\right)^{\frac{1}{2}}
                  \nonumber\\[5pt]&=
                                    \delta_{m,n}^{-1} 
                                    \sqrt{\frac{[N][n]}{[N-m][n+m]}}
                                    \frac{\Gamma(nb^{2}+\nt+1)\Gamma(Nb^{2}+N-1)}{\Gamma((m+n)b^{2}+m+\nt))\Gamma((N-m)(b^{2}+1)))}.
\end{align}
The orthogonal and symmetric S matrix obtained in \eqref{eq:8} is thus
\begin{equation}
\label{eq:smatrix}
  S_{m,n}= \frac{1}{ \sqrt{[m+n][N]}}
  \begin{pmatrix}
    (-1)^{m+1}\sqrt{[N-m][n]}& \sqrt{[N+n][m]} \\[5pt]
    \sqrt{[N+n][m]}&  (-1)^{m}\sqrt{[N-m][n]}
  \end{pmatrix}.
\end{equation}
This is the main result of our paper. For $N=2$ and $n=m=1$, we
recover the Virasoro case \eqref{eq:5}. Therefore, we conclude that
$S \sim S_{m,n}$ up to a normalization redefinition of the conformal
blocks.

Now, to calculate the half-monodromy matrix, let us set $\nt=0$, as
its value only changes monodromy signs and does not change the form of
\eqref{eq:smatrix}. The $s$-channel braiding matrix is obtained from
the asymptotics of the conformal blocks near $z=0$, giving
\begin{equation}
  T_{m,n} =N_{m,n}(q)
  \begin{pmatrix}
    q^{-n}&0\\
    0& (-1)^{m}q^{m}
  \end{pmatrix},
\end{equation}
where we choose a framing in which a $q^{nm/N}$ factor is canceled.
Now, consider the case $n=1$. The two special cases in which we can
construct knots are the parallel case ($m=1$) and the anti-parallel
case ($m=N-1$).  The framing normalization matrices $N_{m,n}(q)$ can
be chosen using the unknot constraint \eqref{eq:knotconstraint} and
Yang-Baxter equation \eqref{eq:yang-baxter}
 \begin{equation}
   N_{1,1}(q) = (-1)^{N+1}q^{2-N}, \quad N_{N-1,1}(q) = 1.
 \end{equation}
 Summing up, the parallel case has the crossing and braiding matrices
\begin{equation}
\label{eq:smatrix2}
  S= \frac{1}{\sqrt{[2][N]}}
  \begin{pmatrix}
    \sqrt{[N-1]}& \sqrt{[N+1]} \\[5pt]
   \sqrt{[N+1]}& - \sqrt{[N-1]}
  \end{pmatrix},\quad   
T =(-1)^{N}
  \begin{pmatrix}
     -q^{1-N}&0\\
    0& q^{3-N}
  \end{pmatrix},
\end{equation}
where $S\equiv S_{1}$, $T\equiv T_{1}$, while the anti-parallel case has
\begin{align}
\label{eq:smatrix3}
  \bar{S}= \frac{1}{[N]}
  \begin{pmatrix}
    1& \sqrt{[N+1][N-1]} \\[5pt]
   \sqrt{[N+1][N-1]}& - 1
 \end{pmatrix},\quad
  \bar{T} =
  \begin{pmatrix}
    q^{-1}&0\\
    0&(-1)^{N+1}q^{N-1}
  \end{pmatrix},
\end{align}
where $\bar{S} \equiv S_{N-1}$ and $\bar{T} \equiv T_{N-1}$. For the
case of braiding two
parallel strands, we can calculate the following invariants
\begin{multline}
  ST^{p}S =\\
\frac{(-1)^{p N}q^{p(2-N)}}{[2][N]}
\left(
\begin{array}{cc}
 (-1)^p [N-1] q^{-p}+ [N+1] q^{p} &  \left((-1)^pq^{-p }-q^{ p}\right) \sqrt{[N-1][N+1]} \\
  \left((-1)^pq^{-p }-q^{ p}\right) \sqrt{[N-1][N+1]} & (-1)^p [N+1] q^{-p}+[N-1] q^{p} \\
\end{array}
\right), 
\end{multline}
which reduces to \eqref{eq:20} when $N=2$. Therefore, we can find, for
example, the knot polynomial for the trefoil knot $3_{1}$, up to framing redefinition,
\begin{equation}
  H_{F}^{3_{1}}(q,A) = \frac{(ST^{3}S)_{22}}{(STS)_{22}} = q^{4}(-1+A^{-2}(q^{2}+q^{-2})).
\end{equation}

We can now try to compare our results with \cite{Zodinmawia:2011oya}
for linking matrices with one link in the fundamental and the other in
an arbitrary symmetric ($m=1$, $n$ arbitrary) or antisymmetric
representation ($m$ arbritrary, $n=1$). For appropriate comparison, we
note that the quantum dimension of a representation $R_{\lambda}$ with
partition $\lambda$ is given by
\begin{equation}
  \text{dim}_{q}R_{\lambda} = \prod_{(i,j)\in \lambda}\frac{[N+j-i]}{[\ell_{i}-i+\ell^{\vee}_{j}-j+1]},
\end{equation}
where $\ell^{\vee}_{j}$ is the number of boxes in the $j$-th column of
$\lambda$. For the antisymmetric case, we get
\begin{equation}
  S_{m,1} = \frac{1}{ \sqrt{[m+1][N]}}
  \begin{pmatrix}
    (-1)^{m+1}\sqrt{[N-m]}& \sqrt{[N+1][m]} \\[5pt]
    \sqrt{[N+1][m]}&  (-1)^{m}\sqrt{[N-m]}
  \end{pmatrix}
\end{equation}
and
\begin{equation}
   S_{N-m,1}= \frac{1}{\sqrt{[N-m+1][N]}}
  \begin{pmatrix}
    (-1)^{N-m+1}\sqrt{[m]}& \sqrt{[N+1][N-m]} \\[5pt]
    \sqrt{[N+1][N-m]}&  (-1)^{N-m}\sqrt{[m]}
  \end{pmatrix}.
\end{equation}
These match, up to signs and column permutation, the first and third
matrices of sec. 4, item 5 of \cite{Zodinmawia:2011oya}. To obtain the
second matrix, necessary to check the Yang-Baxter equation for links,
we have to rederive \eqref{hg-eq} with the fields at $z$ and at $z=0$
interchanged.

For the symmetric case, we get the first and third matrix of sec. 4,
item 3 of \cite{Zodinmawia:2011oya}, up to signs and column
permutation,
\begin{equation}
   S_{1,n}= \frac{1}{\sqrt{[n+1][N]}}
  \begin{pmatrix}
    \sqrt{[N-1][n]}& \sqrt{[N+n]} \\[5pt]
    \sqrt{[N+n]}&  -\sqrt{[N-1][n]}
  \end{pmatrix},
\end{equation}
and
\begin{equation}
   S_{N-1,n}= \frac{1}{ \sqrt{[N][N+n-1]}}
  \begin{pmatrix}
    (-1)^{N}\sqrt{[n]}& \sqrt{[N+n][N-1]} \\[5pt]
    \sqrt{[N+n][N-1]}&  (-1)^{N-1}\sqrt{[n]}
  \end{pmatrix}.
\end{equation}
This strongly suggests that \eqref{eq:smatrix} is the correct crossing
matrix for links with one fundamental component and a rectangular
component. To calculate the $T$ matrix in the correct framing, we need
the other type of matrix mentioned in \cite{Zodinmawia:2011oya} to
apply the constraints.

\section{Conclusions}
\label{sec:conclusions}

In this work, we have obtained crossing and braiding matrices for
certain $W_{N}$ algebra representations. $W_{N}$ primaries are
labelled by two copies of $SU(N)$ representations and, for the type of
correlators described in \cite{Fateev2007a}, we did \emph{not}
discover any new type of Wilson loop invariants apart from the
$SU(N)_{q}$ ones. In particular, we can use these matrices to obtain
HOMFLY knot invariants in the fundamental representation and
two-component HOMFLY link invariants, one component in the fundamental
and another in a rectangular representation of $\sun$ algebra. To
construct generic link invariants in this case, we need three types of
matrices \cite{Zodinmawia:2011oya}, but we have explicitly studied two
types, linking and anti-linking. The third (mixed) case can be
obtained by making a rederivation of \eqref{hg-eq} with the position
of the relevant fields exchanged.

Links with one fundamental component and the other component in an
arbitrary $W_{N}$ representation are related to the generalized
hypergeometric function described in this paper. The connection
problem in this case is more intricate but the monodromy group is well
known \cite{Beukers1989}. This is probably enough to find link
invariants with one fundamental component linked to an arbitrary
higher-representation component, but to get the particular crossing
matrices can be more tricky. In principle, the problem of non-trivial
multiplicity should be automatically solved by the monodromy properties
of the generalized hypergeometric functions. We plan to pursue this
approach in a future work.

The case of knot invariants for higher-representations still remains
elusive. This is due to the limited knowledge we have about
correlation functions in $W_{N}$ models, apart from the cases
discussed in \cite{Fateev2007a,V.A.Fateev:2008aa}. It is known that
correlators more general than the one with a semi-degenerate field do
not obey a linear differential
equation~\cite{Fateev:2005aa,Fateev2007a}. Something can be said about
integral representations of more general correlators
\cite{V.A.Fateev:2008aa}, although via complicated integrals and the
monodromy analysis might be interesting for those results. The
pentagon identity can, in principle, be used to obtain crossing
matrices for knots in higher-representations
\cite{Zodinmawia:2011oya,Gu2014b,D.Galakhov:2015aa}. However, this
approach is computationally expensive from our current
knowledge. Promising results have been recently obtained for Toda
3-point functions and $W_{4}$ crossing matrices in \cite{Furlan2015}
with one of the fields in the representation $\alpha = -b\omega_{2}$
and the other fields in partially degenerate representations
$\beta_{a} =k_{a}\omega_{2}b $. Finally, another interesting approach
to understand $W_{N}$ conformal blocks is via the AGT expansion of
isomonodromic tau-functions
\cite{Gamayun2013a,Iorgov2015,Gavrylenko2015}.  All of this are
relevant lines of attack for the problem of finding Wilson loop
invariants for $W_{N}$ models. We expect that our results will serve
as a basis to further expand the understanding of $W_{N}$ models and
higher-spin topological invariants in Chern-Simons theory.

\acknowledgments
The authors thank Dmitry Melnikov, Andrei Mironov and Alexei Morozov
for important discussions and valuable suggestions. The work of FN was
supported by the CNPq PDJ fellowship of the Science without Borders
initiative process 400635/2012-7.

\bibliographystyle{JHEP} 
\bibliography{knots}

\end{document}